\documentclass[journal]{IEEEtran}

\usepackage[dvips]{graphicx}
\usepackage{graphicx}
\usepackage{amssymb}
\usepackage{cite}
 \usepackage{subcaption}
\usepackage{amsmath}
\usepackage{algorithm}
\usepackage{algpseudocode}
\usepackage{multirow}
\usepackage{xcolor}
\usepackage{enumitem}
\usepackage{amsthm}
\usepackage{threeparttable}
\usepackage{soul}
\usepackage{float}
\usepackage[caption=false]{subfig}
\graphicspath{ {./pic/} }
\begin{document}


\title{Mixture-of-Experts Diffusion Models for Adaptive Massive MIMO Channel Estimation via Variational Bayesian Inference}
\author{Zhuorui Jiang, Jun Fang, Boyu Ning, Hongbin Li,~\IEEEmembership{Fellow,~IEEE}, and Ying-Chang Liang,~\IEEEmembership{Fellow,~IEEE}
\thanks{Zhuorui Jiang, Jun Fang, and Boyu Ning are with the National Key
Laboratory of Wireless Communications, University of
Electronic Science and Technology of China, Chengdu 611731, China,
Email: zrjiang@std.uestc.edu.cn; JunFang@uestc.edu.cn; boydning@outlook.com}
\thanks{Hongbin Li is
with the Department of Electrical and Computer Engineering,
Stevens Institute of Technology, Hoboken, NJ 07030, USA, E-mail:
Hongbin.Li@stevens.edu}
\thanks{Ying-Chang Liang is with Center for Intelligent Networking and Communications, University of Electronic Science and
Technology of China, Chengdu 611731, China, Email: liangyc@ieee.org}
}

\maketitle

\begin{abstract}
Channel estimation is essential to massive multiple-input multiple-output (MIMO) systems.
While recent generative model-based approaches using lightweight diffusion models (DMs) have achieved superior performance, they typically rely on a single data-driven prior, which limits their adaptability to varying channel distributions in real-world scenarios.
To address this deficiency, we propose a mixture-of-experts (MoE) diffusion model (DM) framework combined with variational Bayesian inference.
Specifically, our approach employs multiple pre-trained DMs, with each trained on a specific type of propagation channels.
We then propose a probabilistic graphical model in which the channel is modeled as a latent variable drawn from one of these candidate generative priors with a certain probability. By integrating variational Bayesian inference with DM-based data priors, the underlying channel along with the expert indicator variable are jointly inferred, thus enabling automatic model adaptation for channel estimation. 
The effectiveness of our approach is evaluated on 3GPP CDL channels.
Simulation results demonstrate that our proposed approach achieves a clear performance improvement over the standard DM-based method that employs a single prior trained on aggregated data from all channel types, particularly when the channel samples from different propagation environments are imbalanced. 
\end{abstract}


\section{Introduction}
\IEEEPARstart{M}{assive} multiple-input multiple-output (MIMO) technology has emerged as a cornerstone for 5G and future 6G wireless networks\cite{elijah2022intelligent, wang2024tutorial,UM-MIMO}. 
By deploying a large-scale antenna array at the base station (BS), massive MIMO systems can notably enhance spectral efficiency and link reliability through spatial multiplexing and beamforming\cite{hoseini2017massive,sun2014mimo,kebede2022precoding}. 
However, exploiting the full potential of these benefits relies on the availability of accurate channel state information (CSI)\cite{3gppCSI}. 
Without precise CSI, the system may fail to perform effective precoding and signal detection, inevitably leading to severe performance degradation \cite{guo2022overview}.
Channel estimation typically requires an amount of pilot overhead that scales linearly with the number of transmit antennas \cite{rahman2024deep}.
For massive MIMO systems, such a requirement becomes unacceptable, as it consumes a substantial portion of time-frequency resources and drastically reduces spectral efficiency \cite{raja2023improved,khan2024effective}. 
Consequently, considerable research efforts have been devoted to recovering high-dimensional CSI with a limited training overhead.

\subsection{Related Work}
Traditional linear channel estimators, such as least squares (LS), are widely adopted in practical systems due to their low computational complexity.
However, these methods generally require the number of pilot symbols to be at least equal to the number of transmit antennas and perform poorly in  pilot-deficient scenarios, rendering them unsuitable in massive MIMO scenarios\cite{zhang2022hierarchical}.
To address this limitation, compressed sensing (CS) techniques were employed to exploit sparse or low-rank structures inherent in wireless channels, enabling accurate recovery of the high-dimensional channel with only a small amount of pilot overhead\cite{tibshirani1996regression,sharifi2023channel,donoho2009message,shi2022triple,vila2013expectation,fang2016two,ZhouFang17}.
The practical efficacy of CS-based methods, however, is fundamentally limited by the sparsity assumption imposed on the wireless channel, an assumption which may not always hold in practice.
For instance, channels in rich-scattering urban areas often exhibit non-sparse patterns due to the abundance of multipath components.
In such a case, CS-based methods may incur a severe performance loss\cite{gomez2019sparse}.
Additionally, sparsity is a relatively simplistic, hand-crafted prior that fails to fully capture the intricate structural information including the spatial correlations, clustered multipath propagation, and environment-dependent geometric features inherent in real-world channels\cite{fesl2024channel}.

The above limitation of existing CS-based methods has motivated researchers to explore a radically different data-driven paradigm. 
In particular, with the advent of deep learning (DL), data-driven methods have arisen as a powerful technique demonstrating superior performance\cite{he2018deep,mashhadi2021pruning,dong2019deep}.
These approaches are enabled by diverse data sources, including high-fidelity ray-tracing simulations\cite{ademaj20163gpp}, standardized statistical channel models\cite{sun2017novel}, and real-world measurements collected via experimental testbeds and prototype terminals\cite{chan2019open}.
Supervised DL-based methods typically treat channel estimation as an end-to-end regression task, using neural networks to implicitly learn complex channel features.
For instance, the learned denoising-based approximate message passing (LDAMP) algorithm \cite{he2018deep} replaces the linear shrinkage function in the AMP algorithm with a learnable convolutional neural network (CNN) denoiser, thereby integrating AMP's inference structure with the representational capability of DL.
Similarly, \cite{mashhadi2021pruning} proposes a DL-based framework for joint pilot design and channel estimation, which employs an attention module to capture long-range channel correlations and uses magnitude-based pruning to reduce pilot overhead.
Although these methods achieve competitive performance, they often require extensive labeled data for training and lack generalization to varying pilot configurations or signal-to-noise ratios (SNRs), since they are tailored to specific system setups.

Recognizing these drawbacks, recent research has shifted from supervised learning towards generative models \cite{mirzaei2022deep}.
Generative model-based estimators are designed to capture the underlying distribution of wireless channels in an unsupervised manner.
Early works have employed variational autoencoders \cite{baur2022variational}, generative adversarial networks (GANs) \cite{balevi2021high}, and Gaussian mixture models \cite{su2019channel} to learn expressive channel priors for robust channel estimation.
More recently, diffusion models (DMs) have emerged as a state-of-the-art generative paradigm and attracted significant attention\cite{ho2020denoising}. 
DMs learn to reverse a gradual noising process through iterative denoising steps, demonstrating superior ability in accurately modeling intricate data distributions.
Specifically, the training process of a DM inherently yields an estimate of the score function, which serves as a powerful prior to tackle linear inverse problems such as channel estimation\cite{luo2022understanding,song2020score}. 
With the aid of this learned prior, existing DM-based methods typically perform channel estimation within a posterior sampling framework\cite{arvinte2022mimo,chen2025generative,zhou2025generative}.
One representative approach adopts annealed Langevin dynamics, which iteratively refines estimates by following the score gradient while injecting stochastic noise\cite{arvinte2022mimo}.
Although achieving competitive estimation accuracy, the method incurs high computational cost and latency due to its large number of iterations and substantial model parameters\cite{fesl2024diffusion}.
To enhance efficiency, \cite{zhou2025generative} proposes a deterministic posterior sampling strategy that removes noise perturbation and adopts a lightweight network architecture to reduce inference latency.

\subsection{Contributions of Our Work}
Despite the promise of DM-based estimators, they fundamentally rely on a single generative prior trained on a fixed dataset and assume the training distribution matches all test conditions.
This assumption is often unrealistic in practical scenarios, where users may traverse diverse propagation environments (e.g., urban, suburban and rural areas), causing the ground-truth channel distribution to deviate from the training one \cite{wang2022time}.
While training a universal DM based on aggregated datasets from different channel environments is a straightforward approach, it tends to blur the distinct structural features of each channel type, leading to suboptimal estimation.
Moreover, the amount of data collected from different propagation environments are inherently imbalanced, as the cost and feasibility of data acquisition vary across different propagation scenarios. 
Such imbalance can further dilute the distinct features of data-scarce channel types and mask their unique characteristics.
To address this challenge, we, in this paper, propose a mixture-of-experts (MoE) diffusion framework embedded within a hierarchical probabilistic graphical model, where the channel is treated as a hidden variable generated from one of several candidate propagation-specific priors.
Since conventional MoE approaches based on neural networks for hard expert selection often lack generalization to varying system configurations, we further devise a variational Bayesian inference procedure to jointly estimate the channel and weights of latent experts, enabling fully automatic model adaptation to the real environment.
The main contributions of this work are summarized as follows:

\begin{itemize}
	\item[$\bullet$] 
	A novel MoE diffusion framework is proposed for MIMO channel estimation to address model mismatch in varying environments. Unlike conventional DM-based methods that train a single universal model on aggregated channel data, this framework categorizes the propagation environment into a finite set of types, with each type corresponding to an individual distribution. It employs multiple pre-trained DMs as expert priors, each tailored to capture  a specific distribution, thereby enabling a more accurate prior for posterior inference by adaptively assigning dedicated experts.	
	
	\item[$\bullet$] A DM-empowered variational Bayesian inference algorithm is developed for joint expert selection and channel estimation. The proposed algorithm operates in a fully data-driven manner, relying solely on the received measurements for inference.
	
	\item[$\bullet$] Extensive simulations are conducted on 3GPP Clustered Delay Line (CDL) channel models. Results demonstrate that our approach significantly outperforms all benchmarks across different SNRs and pilot configurations. We also evaluate the complexity of the proposed algorithm. Owing to the fast convergence of the variational inference and the lightweight architecture of the DMs, our proposed method achieves competitive estimation accuracy with low latency.
\end{itemize}

The rest of this paper is organized as follows: Section~\ref{sec:problem formulation} discusses	the system model and the channel estimation problem. 
Section~\ref{sec:DM overview} briefly reviews the fundamentals of DMs and the standard DM-based channel estimation method.
Section~\ref{sec:proposed method} proposes an MoE-based probabilistic graphical model and develops a variational Bayesian inference algorithm for joint expert selection and channel estimation. Simulation results are presented in Section~\ref{sec:simulation}, followed by concluding remarks in Section~\ref{sec:conclusion}.

\emph{Notations:} boldface letters $\mathbf{a}$ and $\mathbf{A}$ denote column vectors and matrices, respectively. 
The superscripts $(\cdot)^T$ and $(\cdot)^H$ represent the transpose and conjugate transpose operations, respectively. 
$p(\mathbf{A})$ stands for the probability density function of a random variable $\mathbf{A}$ and $\nabla_{\mathbf{A}}$ represents the gradient operator with respect to $\mathbf{A}$.
The Frobenius norm and element-wise $l_1$-norm of a matrix are denoted by $\|\cdot\|_{\mathrm{F}}$ and $\|\cdot\|_{\mathrm{m}_1}$, respectively.
The operator $\mathbb{E}[\cdot]$ stands for the expectation, and $\mathcal{CN}(\mathbf{x}; \boldsymbol{\mu}, \boldsymbol{\Sigma})$ denotes the circularly symmetric complex Gaussian distribution with mean $\boldsymbol{\mu}$ and covariance matrix $\boldsymbol{\Sigma}$.

\section{Problem Formulation} \label{sec:problem formulation}
We consider a downlink channel estimation in a large-scale MIMO communication system, where the BS and the user are equipped with $N_t$ and $N_r$ antennas, respectively. The wireless channel is denoted as $\mathbf{H} \in \mathbb{C}^{N_r \times N_t}$. 
To facilitate channel estimation, the BS transmits a sequence of $N_p$ pilot symbols $\mathbf{P}=\left[\mathbf{p}_1\phantom{0} \mathbf{p}_2\phantom{0} \cdots\phantom{0} \mathbf{p}_{N_{p}}\right]\in \mathbb{C}^{N_t \times N_p}$. 
The signal received by the user can be expressed as
\begin{equation} \label{eq:received_signal}
	\mathbf{Y} = \mathbf{H}\mathbf{P} + \mathbf{N},
\end{equation}
where $\mathbf{Y} \in \mathbb{C}^{N_r \times N_p}$, and $\mathbf{N} = \left[\mathbf{n}_1\phantom{0} \mathbf{n}_2\phantom{0} \cdots\phantom{0} \mathbf{n}_{N_{p}}\right] \in \mathbb{C}^{N_r \times N_p}$ is the additive white Gaussian noise (AWGN) with $\mathbf{n}_{k}$ following $\mathcal{CN}(\mathbf{0}, \sigma_{n}^2\mathbf{I})$. 

The objective of channel estimation is to recover the channel $\mathbf{H}$ from the received signal $\mathbf{Y}$. 
Such a problem can be easily solved via a LS-based estimator, provided that the number of pilot symbols is no smaller than the number of transmit antennas, i.e., $N_p\geq N_t$. Nevertheless, for large-scale systems where the BS is equipped with a large number of antenna elements, this would entail a substantial amount of training overhead\cite{rahman2024deep}. 
In this paper, we address this challenge by employing a data-driven approach that leverages a DM to learn the underlying channel distribution. By doing so, we obtain a powerful prior that enables the accurate recovery of high-dimensional channels from a limited number of pilots.

\section{Overview of Generative Diffusion Model-Based Channel Estimation} \label{sec:DM overview}
Developed within a probabilistic generative framework, DMs can approximate complex distributions by learning to reverse a gradual noising process\cite{ho2020denoising}.
In the context of channel estimation, DMs acquire the distribution of the channel $p(\mathbf{H})$ by learning a score function $\nabla_{\mathbf{H}}\log p(\mathbf{H})$, which is the gradient of 
the log-probability density with respect to the channel parameters\cite{luo2022understanding,ho2020denoising}.  
This learned score is pivotal for reconstructing channels from noisy observations, as it effectively guides the iterative inference processes.
Previous DM-based methods either rely on stochastic annealed Langevin dynamics \cite{arvinte2022mimo}, or employ a deterministic posterior sampling technique to infer the posterior channel estimate\cite{zhou2025generative}. 
In the following, we first introduce the mathematical formulation of DMs and then discuss the posterior sampling strategy used to recover the channel.

\subsection{Forward Diffusion Process}
The forward process is defined as a fixed Markov chain that progressively corrupts the data sample $\mathbf{H}_0 \sim p(\mathbf{H}_0)$ by adding Gaussian noise over $T$ discrete time steps. 
Let $\mathbf{H}_t$ represent the latent variable at time step $t$ ($t=1, \dots, T$). 
The transition probability $p(\mathbf{H}_t | \mathbf{H}_{t-1})$ is modeled as a Gaussian distribution, i.e.,
\begin{equation}
	p(\mathbf{H}_t | \mathbf{H}_{t-1}) = \mathcal{CN}(\mathbf{H}_t; \sqrt{1-\beta_t}\mathbf{H}_{t-1}, \beta_t \mathbf{I}),
\end{equation}
where $\{\beta_t\}_{t=1}^T$ are predetermined coefficients satisfying $0 < \beta_1 < \dots < \beta_T < 1$. 
Define $\alpha_t = 1 - \beta_t$ and $\bar{\alpha}_t = \prod_{i=1}^t \alpha_i$.
A notable property of this process is that the marginal distribution $p(\mathbf{H}_t | \mathbf{H}_0)$ at an arbitrary time step $t$ can be expressed in a closed form as
\begin{equation}
	p(\mathbf{H}_t | \mathbf{H}_0) = \mathcal{CN}(\mathbf{H}_t; \sqrt{\bar{\alpha}_t}\mathbf{H}_0, (1 - \bar{\alpha}_t)\mathbf{I}).
\end{equation}
Consequently, $\mathbf{H}_t$ can be directly sampled from $\mathbf{H}_0$ via the reparameterization trick, i.e., $\mathbf{H}_t = \sqrt{\bar{\alpha}_t}\mathbf{H}_0 + \sqrt{1 - \bar{\alpha}_t}\boldsymbol{\epsilon}_t$, where $\boldsymbol{\epsilon}_t \sim \mathcal{CN}(\mathbf{0}, \mathbf{I})$. 
As $t \to T$, the distribution of $\mathbf{H}_T$ eventually converges to a standard isotropic Gaussian distribution, i.e., $\mathcal{CN}(\mathbf{0}, \mathbf{I})$.

\subsection{Reverse Denoising Process}
The generative process is correspondingly modeled as a reverse Markov chain that reconstructs $\mathbf{H}_0$ from the noise prior $\mathbf{H}_T \sim \mathcal{CN}(\mathbf{0}, \mathbf{I})$. 
While the ground-truth reverse posterior $p(\mathbf{H}_{t-1} | \mathbf{H}_t)$ is computationally intractable, it can be obtained when further conditioned on $\mathbf{H}_0$ \cite{ho2020denoising}
\begin{equation}
	p(\mathbf{H}_{t-1} \mid \mathbf{H}_t, \mathbf{H}_0) = \mathcal{CN}\left( \mathbf{H}_{t-1}; \, \tilde{\boldsymbol{\mu}}_t, \, \tilde{\sigma}_t^2 \mathbf{I} \right),
\end{equation}
where $\tilde{\boldsymbol{\mu}}_t = \frac{1}{\sqrt{\alpha_t}} \left( \mathbf{H}_t - \frac{1 - \alpha_t}{\sqrt{1 - \bar{\alpha}_t}} \boldsymbol{\epsilon}_t \right)$ and $\tilde{\sigma}_t^2 = \frac{1 - \bar{\alpha}_{t-1}}{1 - \bar{\alpha}_t} (1-\alpha_t)$. 
To approximate $p(\mathbf{H}_{t-1} | \mathbf{H}_t, \mathbf{H}_0)$, a parameterized distribution $p_{\boldsymbol{\theta}}(\mathbf{H}_{t-1} | \mathbf{H}_t)$ with learnable parameters $\boldsymbol{\theta}$ is 
\begin{equation}
	p_\theta(\mathbf{H}_{t-1} | \mathbf{H}_t) = \mathcal{CN}(\mathbf{H}_{t-1}; \boldsymbol{\mu}_\theta(\mathbf{H}_t, t),\tilde{\sigma}_t^2 \mathbf{I}),
\end{equation}
where $\boldsymbol{\mu}_{\boldsymbol{\theta}}(\mathbf{H}_t, t)$ stands for the predicted mean.
To approximate the true posterior mean $\tilde{\boldsymbol{\mu}}_t$, the mean $\boldsymbol{\mu}_{\boldsymbol{\theta}}$ is set as
\begin{equation}
	\boldsymbol{\mu}_{\boldsymbol{\theta}}(\mathbf{H}_t, t) = \frac{1}{\sqrt{\alpha_t}} \left( \mathbf{H}_t - \frac{1-\alpha_t}{\sqrt{1 - \bar{\alpha}_t}} \boldsymbol{\epsilon_\theta}(\mathbf{H}_t, t) \right).
\end{equation}

Eventually it can be shown that training a DM boils down to learning a denoising network $\boldsymbol{\epsilon}_{\boldsymbol{\theta}}(\mathbf{H}_t, t)$ to predict the noise $\boldsymbol{\epsilon}_t$ with the current time step $t$ and $\mathbf{H}_{t}$. The loss function based on the true and predicted noise is given by
\begin{equation} \label{eq:DM_training}
	\mathcal{L}(\boldsymbol{\theta}) = \mathbb{E}_{t,\, \mathbf{H}_0,\, \boldsymbol{\epsilon}_t} \left[ \left\| \boldsymbol{\epsilon}_t - \boldsymbol{\epsilon}_{\boldsymbol{\theta}}\!\left( \sqrt{\bar{\alpha}_t}\,\mathbf{H}_0 + \sqrt{1 - \bar{\alpha}_t}\,\boldsymbol{\epsilon}_t,\, t \right) \right\|^2 \right],
\end{equation}
where $t$ is sampled uniformly at random from $\{1, 2, \dots, T\}$. 
It is worth noting that the training process is unsupervised, since it does not require any knowledge of the pilot matrix $\mathbf{P}$ or the received signal $\mathbf{Y}$.
After training, $\boldsymbol{\theta}$ is determined, and the data sample can be generated following the iterative process from $\mathbf{H}_{T}$ to $\mathbf{H}_{0}$ as
\begin{equation} \label{eq:DM_reverse}
	\mathbf{H}_{t-1} = \frac{1}{\sqrt{\alpha_t}} \left( \mathbf{H}_t - \frac{1 - \alpha_t}{\sqrt{1 - \bar{\alpha}_t}} \boldsymbol{\epsilon_\theta}(\mathbf{H}_t, t) \right) + \tilde{\sigma}_t \mathbf{Z}_t
\end{equation}
where $\mathbf{Z}_{t} \sim \mathcal{CN}(\mathbf{0}, \mathbf{I})$ is the random noise perturbation.

Interestingly, based on Tweedie’s formula \cite{efron2011tweedie}, the trained noise prediction network $\boldsymbol{\epsilon_\theta}(\mathbf{H}_t, t)$ can serve as an estimator of the score function. As shown in \cite{luo2022understanding}, we have
\begin{equation} \label{eq:prior_score_function}
	\nabla_{\mathbf{H}_t} \log p(\mathbf{H}_t) \approx -\frac{1}{\sqrt{1 - \bar{\alpha}_t}} \boldsymbol{\epsilon}_{\boldsymbol{\theta}}(\mathbf{H}_t, t).
\end{equation}
This formulation enables the pre-trained DM to provide an approximation of the score function of the channel distribution, which can be used for posterior inference.

\subsection{DM-Based Channel Estimation}
Since the pre-trained DM directly provides the prior score $\nabla_{\mathbf{H}_{t}} \log p(\mathbf{H}_{t})$, the channel estimation task can be formulated as a posterior sampling problem. 
The objective is to generate samples from the posterior distribution $p(\mathbf{H}|\mathbf{Y})$ given the received $\mathbf{Y}$. 
To this end, Bayes' rule is leveraged to decompose the intractable posterior score into two distinct components as\cite{meng2022diffusion,zhou2025generative}
\begin{equation}
	\nabla_{\mathbf{H}_t} \log p(\mathbf{H}_t|\mathbf{Y}) = \nabla_{\mathbf{H}_t} \log p(\mathbf{Y}|\mathbf{H}_t) + \nabla_{\mathbf{H}_t} \log p(\mathbf{H}_t).
\end{equation} 
Clearly, $\nabla_{\mathbf{H}_t} \log p(\mathbf{H}_t)$ can be obtained from \eqref{eq:prior_score_function} via the denoising network $\boldsymbol{\epsilon_\theta}(\mathbf{H}_t, t)$. Based on how the posterior score is utilized in the iterative recovery process, existing methods can be broadly categorized into stochastic and deterministic posterior sampling methods.

The stochastic approach employs annealed Langevin dynamics to sample from the posterior. 
The process begins with $\mathbf{H}_T$ and progressively anneals the noise level from $t = T$ to 0. 
At each time step $t$, the approach first determines the decaying step-size parameters $\zeta_t$ and $\xi_t$. Multiple iterations are then performed following the direction of the posterior score while simultaneously injecting random noise to avoid local optima\cite{arvinte2022mimo,song2019generative}. 
The update rule is given by
\begin{equation} \label{eq:Annealed_Langevin}
	\mathbf{H}_{t,i+1} =  \mathbf{H}_{t,i} + \zeta_{t} \nabla_{\mathbf{H}_{t,i}} \log p(\mathbf{H}_{t,i}|\mathbf{Y}) + \xi_t \mathbf{Z}_{t,i},
\end{equation}
where $\mathbf{Z}_{t,i} \sim \mathcal{CN}(\mathbf{0}, \mathbf{I})$ is the injected perturbation in the $i$-th step at time step $t$. 
While a robust performance can be achieved by this method, due to the stochastic nature, it typically requires a double loop of iterations to converge, leading to significant inference latency \cite{zhou2025generative,fesl2024diffusion}.

To recover the channel with fewer time steps, a recent work \cite{zhou2025generative} proposed a deterministic posterior sampling update rule as follows
\begin{equation} \label{eq:deterministic_posterior_samling}
	\mathbf{H}_{t-1} = \frac{1}{\sqrt{\alpha_{t}}} \left( \mathbf{H}_{t} + (1 - \alpha_{t}) \nabla_{\mathbf{H}_{t}} \log p(\mathbf{H}_{t}|\mathbf{Y}) \right).
\end{equation}
The update rule can be interpreted as yielding a posterior mean estimate of the latent variable $\mathbf{H}_{t-1}$ conditioned on $\mathbf{H}_{t}$ and $\mathbf{Y}$. 
Similarly, the iteration proceeds from $t = T$, and the channel can be finally recovered as $\mathbf{\hat{H}} = \mathbf{H}_{0}$.

\section{Proposed Method} \label{sec:proposed method}
Existing DM-based channel estimation methods are trained on data samples following a specified distribution, and excel when the test distribution matches the training data. However, in practice, mobile users may travel from one place to another\cite{wang2022time}. 
These different areas traversed by the user usually exhibit diverse propagation environments and thus are characterized by different channel features and distributions. 
Generally, the propagation environments can be categorized into a finite number of types such as urban, suburban and rural areas, with each type corresponding to an individual distribution.  

Ideally, for each type of propagation environment, one can train a dedicated DM using samples collected specifically under this type of channel condition. 
Then, based on the user's current location, one can choose a corresponding pre-trained DM as the prior to perform the posterior channel inference. 
Nevertheless, in practice, the prior knowledge of the user's location and its associated type of channel environment may not be available during inference or may change too rapidly to be identified \cite{arvinte2022mimo}.   
Without this knowledge, a straightforward approach to circumvent the performance degradation caused by distribution mismatch is to aggregate channel samples from all propagation environments and train a single DM. 
Such an approach, however, tends to dilute each type of channel's distinctive structural features and leads to suboptimal estimation.
This effect can be further exacerbated when training datasets from different propagation environments are imbalanced and certain scenarios are severely data-scarce \cite{wang2016training}.

\begin{figure}[!t]
	\centering
	\includegraphics[width=0.95\linewidth]{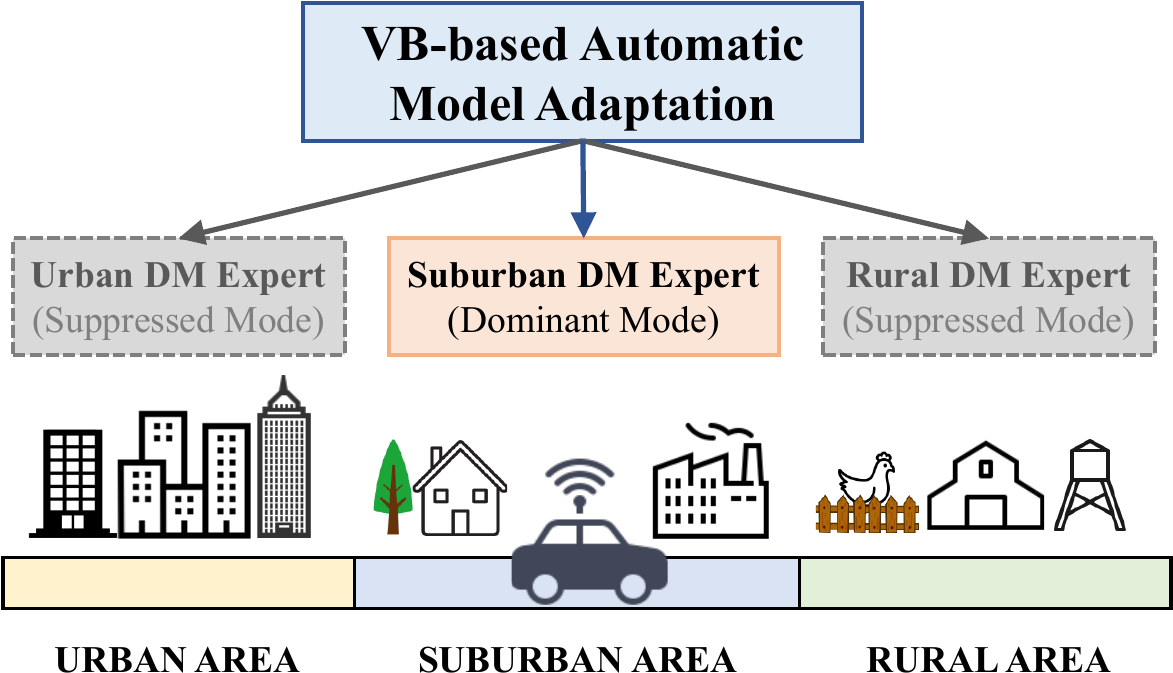}
	\caption{Illustration of the proposed VB-based automatic model adaptation framework.}
	\label{fig:DMVB_framework}
\end{figure}

To address the challenge, this paper proposes a MoE DM-empowered variational Bayes (VB) framework, which integrates variational Bayesian inference with the powerful learning capabilities of generative AI models. 
Specifically, for each type of channel, we train a corresponding DM to capture each type of channel's distribution. 
The channel to be inferred is modeled as a latent variable following a probabilistic mixture model, which assumes the true channel is generated from one of a finite set of pre-trained distributions with a certain probability.  
The variational Bayesian inference algorithm can thus be utilized to identify the latent source distribution based on the observations.
This algorithm enables automatic selection of an appropriate DM to provide a precise prior for posterior channel estimation. 
Fig.~\ref{fig:DMVB_framework} provides a schematic of the proposed VB-based automatic model adaptation framework. 

\subsection{Training of Generative Diffusion Models}
To implement this MoE framework, it is necessary to train each expert DM according to its respective dataset.
Instead of applying a sophisticated architecture, we follow \cite{zhou2025generative} and adopt a lightweight CNN for the denosing network $\boldsymbol{\epsilon_\theta}$ in our work. 
This network directly operates on the spatial channel matrix to capture the intricate local correlations inherent in the spatial domain.

Specifically, the complex-valued noisy channel $\mathbf{H}_{t} \in \mathbb{C}^{N_r \times N_t}$ is decomposed into its real and imaginary components, resulting in a real-valued input tensor of size $2 \times N_r \times N_t$.
The time step $t$ is specified to the network as a vector $\mathbf{t} \in \mathbb{R}^{S_{\mathrm{init}}}$ via Transformer sinusoidal positional embeddings, where $S_{\mathrm{init}}$ denotes the initial time embedding dimension.
This feature vector $\mathbf{t}$ is subsequently passed through a single fully connected dense layer to yield an intermediate representation, which is then split into a scaling vector $\mathbf{t}_s$ and a bias vector $\mathbf{t}_b$, with both sizes equal to $S_{\mathrm{max}}$.

\begin{figure}[!t]
	\centering
	\includegraphics[width=0.98\linewidth]{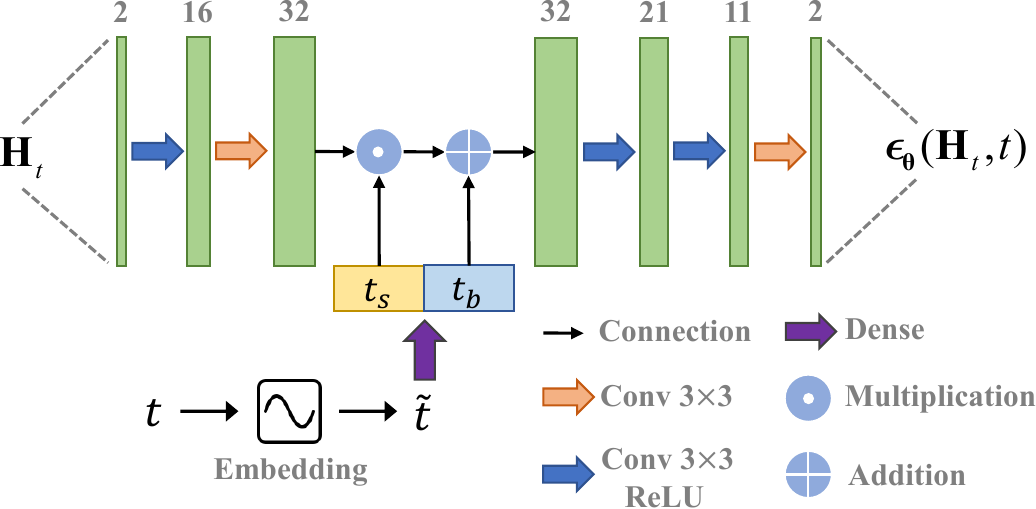}
	\caption{Architecture of the DM's denoising network, based on a lightweight CNN with embedded time-step conditioning.}
	\label{fig:CNN}
\end{figure}

The network is mainly composed of a sequence of 2D convolutional layers with $3 \times 3$ kernels.
To extract high-dimensional spatial features, the architecture first progressively increases the number of kernels in each layer to a maximum feature width of $S_{\mathrm{max}}$ through two convolutional layers, where the first layer employs rectified linear unit (ReLU) activation.
Let $\mathcal{X} \in \mathbb{R}^{ S_{\mathrm{max}} \times N_r \times N_t}$ denote the extracted intermediate feature tensor. To fuse the time-step condition, we apply a channel-wise affine transformation, which operates along each mode-1 fiber (tube) of the tensor. Specifically, for a fixed spatial location $(i, j)$, the fiber $\mathcal{X}_{:ij} \in \mathbb{R}^{S_{\mathrm{max}}}$ is transformed independently as
\begin{equation}
    (\mathcal{X}_{\mathrm{mod}})_{:ij} = (1 + \mathbf{t}_s) \odot \mathcal{X}_{:ij} + \mathbf{t}_b,
\end{equation}
where $\odot$ denotes the Hadamard (element-wise) product. Following this, three convolutional layers gradually reduce the feature depth back to the input size.
The first two layers are equipped with ReLU activation, while the final one outputs the predicted noise $\boldsymbol{\epsilon_\theta} \in \mathbb{R}^{2 \times N_r \times N_t}$. 
The architecture of the DM's denoising network is illustrated in Fig.~\ref{fig:CNN} for clarity. 

At each training iteration, a time index $t \in \{1, 2, \dots, T\}$ is randomly sampled and a clean channel sample $\mathbf{H}_0$ is drawn from the training dataset. 
We corrupt $\mathbf{H}_0$ with random noise $\boldsymbol{\epsilon}_t \sim \mathcal{CN}(\mathbf{0}, \mathbf{I})$, following $\mathbf{H}_t = \sqrt{\bar{\alpha}_t}\mathbf{H}_0 + \sqrt{1 - \bar{\alpha}_t}\boldsymbol{\epsilon}_t$. 
The denoising network, parameterized by $\boldsymbol{\theta}$, predicts the noise $\boldsymbol{\epsilon_\theta}(\mathbf{H}_t, t)$. The parameters $\boldsymbol{\theta}$ then can be updated via back propagation to minimize the discrepancy between the true noise and the predicted one, as defined in \eqref{eq:DM_training}.

\subsection{Probabilistic Graphical Model}
Given a number of pretrained DMs, we introduce the following probabilistic graphical model to learn the unknown channel $\mathbf{H}$ from the received pilot measurements $\mathbf{Y}$.
Specifically, to enable automatic model adaptation, the channel is modeled as a latent variable governed by a probabilistic mixture model. 
Let $\mathbf{z}$ be a one-hot vector such that $z_{k} = 1$ if the $k$-th expert is selected and $z_{j} = 0, j \neq k$ otherwise.  
The conditional distribution of the channel given $\mathbf{z}$ is defined as
\begin{align}
p(\mathbf{H} | \mathbf{z}) \triangleq \prod_{k=1}^K [p_k(\mathbf{H})]^{z_k}
\end{align}
where $p_k(\mathbf{H})$ represents the $k$-th component prior distribution whose score function $\nabla_{\mathbf{H}} \log p_k(\mathbf{H})$ is learned by the $k$-th pre-trained DM with parameters $\boldsymbol{\theta}_k$.

To learn the latent indicator variable $\mathbf{z}$ automatically, it is assumed to follow a categorical distribution parameterized by the mixing coefficients $\boldsymbol{\pi} = [\pi_1, \pi_2, \dots, \pi_K]^T$ as
\begin{align}
p(\mathbf{z} | \boldsymbol{\pi}) = \prod_{k=1}^K \pi_k^{z_k},
\end{align}
where each $\pi_k \in [0, 1]$ denotes the probability of selecting the $k$-th component and we have $\sum_{k=1}^K \pi_k = 1$. 

In addition, the mixing coefficients $\boldsymbol{\pi}$ are treated as random variables following a Dirichlet distribution, as it serves as the conjugate prior to the categorical distribution. The density is given by
\begin{equation}
	p(\boldsymbol{\pi}) = \text{Dir}(\boldsymbol{\pi} \mid \pmb{\gamma}) = \frac{\Gamma(\sum_{k=1}^K \gamma_{k})}{\prod_{k=1}^K \Gamma(\gamma_{k})} \prod_{k=1}^K \pi_k^{\gamma_{k} - 1},
\end{equation}
where $\pmb{\gamma} = [\gamma_{1}, \dots, \gamma_{K}]^T$ is the vector of concentration hyperparameters, and $\Gamma(\cdot)$ denotes the Gamma function.
This Dirichlet prior facilitates the incorporation of prior beliefs regarding the prevalence of different channel types in practical scenarios.
When no specific preference is assumed, setting $\gamma_{k} = 1$ for all $k$ provides an unbiased prior across all experts. 

\begin{figure}[!t]
	\centering
	\includegraphics[width=0.7\linewidth]{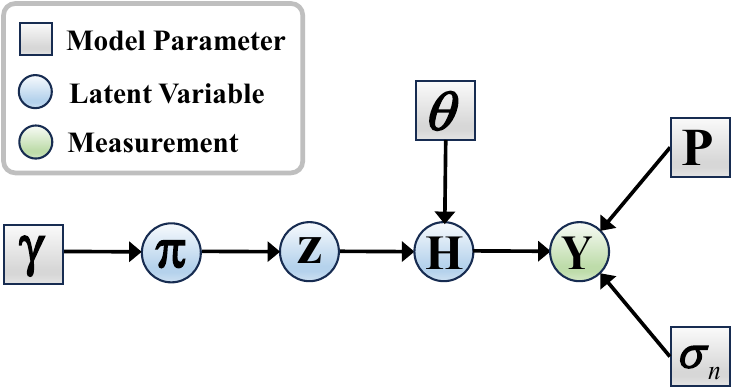}
	\caption{Illustration of the proposed probabilistic graphical model.}
	\label{fig:graphical_model}
\end{figure}

Given the channel estimation model \eqref{eq:received_signal}, it can be observed that the likelihood of the received measurement $\mathbf{Y}$ follows a Gaussian distribution, i.e.,
\begin{align}
p(\mathbf{Y} | \mathbf{H}) = \mathcal{CN}(\mathbf{Y} | \mathbf{H}\mathbf{P}, \sigma_n^2 \mathbf{I})
\end{align}
By combining the likelihood and the hierarchical priors, the joint probability distribution of the received signal and the latent variables can be factored as
\begin{equation} \label{eq:bayes_hierarchy}
	p(\mathbf{Y}, \mathbf{H}, \mathbf{z}, \boldsymbol{\pi}) = p(\mathbf{Y} | \mathbf{H}) p(\mathbf{H} | \mathbf{z}) p(\mathbf{z} | \boldsymbol{\pi}) p(\boldsymbol{\pi}).
\end{equation}
Accordingly, the proposed probabilistic graphical model is depicted in Fig.~\ref{fig:graphical_model}.

The objective of the proposed method is to infer the posterior distribution $p(\mathbf{H}, \mathbf{z}, \boldsymbol{\pi} | \mathbf{Y})$. 
By applying Bayes' rule, it can be explicitly expressed as
\begin{equation}\\
	\scalebox{0.99}{$\displaystyle
		p(\mathbf{H}, \mathbf{z}, \boldsymbol{\pi} | \mathbf{Y}) = \frac{p(\mathbf{Y} | \mathbf{H}) p(\mathbf{H} | \mathbf{z}) p(\mathbf{z} | \boldsymbol{\pi}) p(\boldsymbol{\pi})}{\sum_{\mathbf{z}} \iint p(\mathbf{Y} | \mathbf{H}) p(\mathbf{H} | \mathbf{z}) p(\mathbf{z} | \boldsymbol{\pi}) p(\boldsymbol{\pi}) \, d\mathbf{H} \, d\boldsymbol{\pi}}. $}
\end{equation}

Clearly, the denominator involves high-dimensional integration over the latent space. 
Coupled with the highly complex priors $p_k(\mathbf{H})$ induced by DMs, the exact computation of this posterior is analytically intractable.   
Therefore, a variational Bayesian inference approach is derived in the following to approximate the posterior distributions.

\subsection{Variational Bayesian Inference}
Let $\boldsymbol{\Theta} \triangleq \{\mathbf{H}, \mathbf{z}, \boldsymbol{\pi}\}$ denote the set of latent variables in the probabilistic graphical model.
The objective of Bayesian inference is to
find the posterior distribution of the latent variables given the observed data, i.e., $p(\boldsymbol{\Theta} | \mathbf{Y})$. 
Notice that the log-marginal probability of the observed data $\log p(\mathbf{Y})$ can be decomposed into two terms\cite{wan2017robust}:
\begin{equation}
	\scalebox{0.93}{$\displaystyle
	\log p(\mathbf{Y}) = \underbrace{\int q(\boldsymbol{\Theta}) \log \frac{p(\mathbf{Y}, \boldsymbol{\Theta})}{q(\boldsymbol{\Theta})} d\boldsymbol{\Theta}}_{\mathcal{L}(q)} \ 
	\underbrace{-\int q(\boldsymbol{\Theta}) \log \frac{p(\boldsymbol{\Theta}|\mathbf{Y})}{q(\boldsymbol{\Theta})} d\boldsymbol{\Theta}}_{\mathrm{KL}(q \| p)} $}
\end{equation}
where $q(\boldsymbol{\Theta})$ represents any arbitrary probability density function,  $\mathcal{L}(q) = \mathbb{E}_{q}[\log p(\mathbf{Y}, \boldsymbol{\Theta}) - \log q(\boldsymbol{\Theta})]$ is the Evidence Lower Bound (ELBO) and $\text{KL}(q \| p)$ denotes the Kullback-Leibler (KL) divergence between the approximate distribution $q(\boldsymbol{\Theta})$ and the true posterior $p(\boldsymbol{\Theta} | \mathbf{Y})$.
Considering that the KL divergence is non-negative and $\log p(\mathbf{Y})$ is constant with respect to the variational parameters, $\mathcal{L}(q)$ serves as the lower bound on $\log p(\mathbf{Y})$ and maximizing the ELBO is equivalent to minimizing the KL divergence.

To make the optimization tractable, the mean field theory is utilized \cite{barabasi1999mean}, 
where the latent variables are assumed to be mutually independent, thus permitting the factorization of $q(\boldsymbol{\Theta}) \approx q(\mathbf{H}) q(\mathbf{z}) q(\boldsymbol{\pi})$.
By decomposing the posterior into independent factors, the optimization problem becomes manageable.
Specifically, the problem is formulated as maximizing the ELBO term to this factorized form, i.e.,
$\max_{q(\mathbf{H}), q(\mathbf{z}), q(\boldsymbol{\pi})} \mathcal{L}(q)$.
Hence, the inference task can be performed via alternating updates over each latent variable. 
The general update rule for an arbitrary variational factor $q(\boldsymbol{\Theta}_j)$ can be expressed as \cite{tzikas2008variational}
\begin{equation}
	\log q^*(\boldsymbol{\Theta}_j) = \mathbb{E}_{q_{-j}}[\log p(\mathbf{Y}, \boldsymbol{\Theta})] + \text{const},
\end{equation}
where $\mathbb{E}_{q_{-j}}[\cdot]$ denotes the expectation with respect to all the variational distributions except $q(\boldsymbol{\Theta}_j)$.
Applying this general rule to the proposed hierarchical model, the ELBO
maximization can be conducted in an alternating fashion for each latent variable, which leads to
\begin{subequations}\label{eq:update_all}
	\begin{align}
		\log q^*(\mathbf{H}) &= \mathbb{E}_{q(\mathbf{z})q(\boldsymbol{\pi})}[\log p(\mathbf{Y}, \mathbf{H}, \mathbf{z}, \boldsymbol{\pi})] + \text{const}, \label{eq:update_qh} \\
		\log q^*(\mathbf{z}) &= \mathbb{E}_{q(\mathbf{H})q(\boldsymbol{\pi})}[\log p(\mathbf{Y}, \mathbf{H}, \mathbf{z}, \boldsymbol{\pi})] + \text{const}, \label{eq:update_qz} \\
		\log q^*(\boldsymbol{\pi}) &= \mathbb{E}_{q(\mathbf{H})q(\mathbf{z})}[\log p(\mathbf{Y}, \mathbf{H}, \mathbf{z}, \boldsymbol{\pi})] + \text{const}. \label{eq:update_qpi}
	\end{align}
\end{subequations}

These factors are updated iteratively until convergence, as detailed in the following.

1) \textbf{Update of} $q(\mathbf{H})$: 
Substituting the factorized joint distribution \eqref{eq:bayes_hierarchy} into \eqref{eq:update_qh}, the variational distribution $q(\mathbf{H})$ can be obtained as
\begin{equation} \label{eq:qh_approx}
	\begin{aligned}
		\log q^*(\mathbf{H}) 
		&\propto \mathbb{E}_{q(\mathbf{z})q(\boldsymbol{\pi})}\Big[ 
		\log p(\mathbf{Y} \mid \mathbf{H}) 
		+ \log p(\mathbf{H} \mid \mathbf{z}) \\
		&\quad
		+ \log p(\mathbf{z} \mid \boldsymbol{\pi}) 
		+ \log p(\boldsymbol{\pi}) 
		\Big] \\
		&\propto \log p(\mathbf{Y} \mid \mathbf{H}) 
		+ \sum_{k=1}^K \rho_k \log p_k(\mathbf{H})
	\end{aligned}
\end{equation}
where $\rho_k = q(z_k=1, z_{j \neq k}=0)$ denotes the posterior probability that the $k$-th expert is selected, and $\log p_k(\mathbf{H})$ is the log-prior from the $k$-th pre-trained diffusion model.

The above equation suggests that the approximate posterior of the channel is a trade-off between data consistency and a mixture of expert priors, weighted specifically by
$\boldsymbol{\rho} = [\rho_1\phantom{0} \rho_2\phantom{0} \dots\phantom{0} \rho_K]$.
However, since the channel priors $p_k(\mathbf{H})$ are implicitly defined by the DMs, obtaining the posterior $q^*(\mathbf{H})$ in a closed form is intractable.

To circumvent this obstacle, we resort to the maximum a posteriori (MAP) strategy, which is equivalent to approximating $q(\mathbf{H})$ with a Dirac distribution centered at the mode, i.e., $q(\mathbf{H}) \approx \delta(\mathbf{H} - \hat{\mathbf{H}})$\cite{osgood2019lectures}. 
To search for the MAP estimate of $\mathbf{H}$, we employ the following deterministic 
posterior sampling paradigm developed within the DM framework, i.e.,
\begin{equation} \label{eq:h_update}
		\mathbb{E}\left[\mathbf{H}_{t-1}|\mathbf{H}_t,\mathbf{Y}\right] =  \frac{1}{\sqrt{\alpha_{t}}} \left( \mathbf{H}_{t} + (1-\alpha_{t}) \nabla_{\mathbf{H}_{t}} \log p(\mathbf{H}_{t}|\mathbf{Y}) \right). 
\end{equation}
Here we replace the posterior $p(\mathbf{H}_{t}|\mathbf{Y})$ with its approximation $q^*(\mathbf{H})$.
Thus, $\nabla_{\mathbf{H}_{t}} \log p(\mathbf{H}_{t}|\mathbf{Y})$ can be rewritten as
\begin{equation} \label{eq:MAP_posterior_decomposition}
	\scalebox{0.90}{$\displaystyle
	\begin{aligned}
			\nabla_{\mathbf{H}_{t}} \log p(\mathbf{H}_{t}|\mathbf{Y}) &= \sum_{k=1}^K \rho_k \nabla_{\mathbf{H}_t} \log p_k(\mathbf{H}_{t}) + \nabla_{\mathbf{H}_t} \log p(\mathbf{Y}|\mathbf{H}_t) \\
			&\approx \sum_{k=1}^K \rho_k \nabla_{\mathbf{H}_t} \log p_k(\mathbf{H}_{t}) + s \cdot \nabla_{\mathbf{H}_t} \log p(\mathbf{Y}|\mathbf{H}_t),	
	\end{aligned} $}
\end{equation}
where, in the second line, we introduce a scaling factor $s \geq 1$ to provide a tradeoff between the prior score and the likelihood score. 
According to \cite{zhou2025generative,dhariwal2021diffusion}, a larger value of $s$ gives a higher importance to the likelihood, which improves the mode of the posterior distribution and enhances estimation performance.

By utilizing \eqref{eq:prior_score_function}, the $\rho_k$-weighted sum of the prior scores can be computed as
\begin{equation} \label{eq:prior_mix_score}
	\sum_{k=1}^K \rho_k \nabla_{\mathbf{H}_t} \log p_k(\mathbf{H}_{t}) \approx -\sum_{k=1}^K \frac{\rho_k}{\sqrt{1 - \bar{\alpha}_t}} \boldsymbol{\epsilon}_{\boldsymbol{\theta}_k}(\mathbf{H}_t, t).
\end{equation}
Now we discuss how to calculate the term $\nabla_{\mathbf{H}_t} \log p(\mathbf{Y}|\mathbf{H}_t)$ in (\ref{eq:MAP_posterior_decomposition}). In DMs, 
the measurement $\mathbf{Y}$ is generated from the clean channel $\mathbf{H}_0$ via $\mathbf{Y} = \mathbf{H}_0\mathbf{P} + \mathbf{N}$, and the calculation of $\nabla_{\mathbf{H}_t} \log p(\mathbf{Y}|\mathbf{H}_t)$ requires an intractable marginalization over $\mathbf{H}_0$:
\begin{equation}
	p(\mathbf{Y}|\mathbf{H}_t) = \int p(\mathbf{Y}|\mathbf{H}_0) p(\mathbf{H}_0|\mathbf{H}_t) d\mathbf{H}_0.
\end{equation}
To address this issue, by leveraging the uninformative prior assumption, the work \cite{meng2022diffusion} proposes to approximate $p(\mathbf{H}_0|\mathbf{H}_t) \propto p(\mathbf{H}_t|\mathbf{H}_0) \approx \mathcal{CN}(\mathbf{H}_0; \frac{1}{\sqrt{\bar{\alpha}_t}}\mathbf{H}_t, \frac{1 - \bar{\alpha}_t}{\bar{\alpha}_t}\mathbf{I})$.
Under this approximation, the noise-perturbed likelihood $p(\mathbf{Y}|\mathbf{H}_t)$ takes the form of a Gaussian distribution, as derived in \cite{zhou2025generative}
\begin{equation} \label{eq:p(y|ht)}
	p(\mathbf{Y}|\mathbf{H}_t) = \mathcal{CN}\left( \mathbf{Y}; \frac{1}{\sqrt{\bar{\alpha}_t}}\mathbf{H}_t\mathbf{P}, \frac{1 - \bar{\alpha}_t}{\bar{\alpha}_t}\mathbf{P}^H\mathbf{P} + \sigma_n^2\mathbf{I} \right),
\end{equation} 
and thus the likelihood score is given by
\begin{equation} \label{eq:likelihood_score}
	\scalebox{0.97}{$\displaystyle
		\begin{aligned}
			&\nabla_{\mathbf{H}_t} \log p(\mathbf{Y}|\mathbf{H}_t) \\
			&= \frac{1}{\sqrt{\bar{\alpha}_t}} \left( \mathbf{Y}\mathbf{V} - \frac{1}{\sqrt{\bar{\alpha}_t}} \mathbf{H}_t\mathbf{U}\mathbf{\Sigma} \right) \left( \frac{1 - \bar{\alpha}_t}{\bar{\alpha}_t} \mathbf{\Sigma}^2 + \sigma_n^2 \mathbf{I} \right)^{-1}\mathbf{\Sigma}\mathbf{U}^H,
		\end{aligned}$}
\end{equation}
where the singular value decomposition (SVD), i.e., $\mathbf{P} = \mathbf{U}\boldsymbol{\Sigma}\mathbf{V}^{H}$ is adopted to mitigate the computational cost of the matrix inversion.

By integrating the weighted prior and likelihood scores into the deterministic update rule, the MAP estimate of $\mathbf{H}$ can be obtained via the iterative process (\ref{eq:h_update}) (from $t = T$ to $t=1$).

2) \textbf{Update of} $q(\mathbf{z})$: Similarly, the variational optimization of the latent indicator $q(\mathbf{z})$ yields
\begin{equation} \label{eq:rho_update}
	\begin{aligned}
		\log q^*(\mathbf{z}) &\propto \mathbb{E}_{q(\mathbf{H})q(\boldsymbol{\pi})}[\log p(\mathbf{H} | \mathbf{z}) + \log p(\mathbf{z} | \boldsymbol{\pi})] \\
		&\propto \sum_{k=1}^K z_k \left( \mathbb{E}_{q(\mathbf{H})}[\log p_k(\mathbf{H})] + \mathbb{E}_{q(\boldsymbol{\pi})}[\log \pi_k] \right).
	\end{aligned}
\end{equation}
Notably, this form identifies $q^*(\mathbf{z})$ as a categorical distribution, i.e.,  $q^*(\mathbf{z}) = \prod_{k=1}^K \rho_k^{z_k}$, where 
\begin{equation}
	\log \rho_k = \mathbb{E}_{q(\mathbf{H})}[\log p_k(\mathbf{H})] + \mathbb{E}_{q(\boldsymbol{\pi})}[\log \pi_k].
\end{equation}

Hence, $q^*(\mathbf{z})$ can be obtained by updating $\boldsymbol{\rho}$.
It is clear that $\log\rho_k$ is a sum of two terms. 
Premised on the MAP estimate $\hat{\mathbf{H}}$, the first term reduces to $\log p_{k}(\hat{\mathbf{H}})$.
While $\log p_{k}(\hat{\mathbf{H}})$ is intractable, approximating it with its ELBO is viable. 
As rigorously derived in \cite{luo2022understanding}, $\log p_{k}(\hat{\mathbf{H}})$ is lower-bounded by
\begin{equation}
	\scalebox{0.99}{$\displaystyle
	\begin{aligned} \label{eq:ELBO_approx}
		&\log p_k(\hat{\mathbf{H}}) \ge \mathcal{L}^{(k)}(\hat{\mathbf{H}}) \\ 
		&\approx -\sum_{t=1}^T \mathbb{E}_{p_k(\mathbf{H}_t|\hat{\mathbf{H}})} \left[ D_{\text{KL}}(p_k(\mathbf{H}_{t-1}|\mathbf{H}_t, \hat{\mathbf{H}}) \| p_{\boldsymbol{\theta}_k}(\mathbf{H}_{t-1}|\mathbf{H}_t)) \right] \\
		&\approx -\sum_{t=1}^T  \mathbb{E}_{\boldsymbol{\epsilon}_t} \left[ \|\boldsymbol{\epsilon}_t - \boldsymbol{\epsilon}_{\boldsymbol{\theta}_k}(\sqrt{\bar{\alpha}_t}\hat{\mathbf{H}} + \sqrt{1-\bar{\alpha}_t}\boldsymbol{\epsilon}_t, t)\|^2 \right],
	\end{aligned} $}
\end{equation}
To reduce the computational complexity, in practice, the sum from $t=1$ to $T$ in \eqref{eq:ELBO_approx} can be simplified by subsampling time steps within this range and then summing these subsampled points. 
For each selected $t$, the expectation is estimated via Monte Carlo sampling, with typically just one draw. Let $\mathcal{T}_{\text{sub}}$ denote the set of selected time steps and $\boldsymbol{\epsilon}_t^{(m)} \sim \mathcal{CN}(\mathbf{0}, \mathbf{I})$ denote the noise sample, we thus have
\begin{equation} \label{eq:prior_approx}
	\scalebox{0.94}{$\displaystyle
		\log p_k(\hat{\mathbf{H}}) \approx -\sum_{t \in \mathcal{T}_{\text{sub}}} \left\| \boldsymbol{\epsilon}_t^{(m)} - \boldsymbol{\epsilon}_{\boldsymbol{\theta}_k}\big( \sqrt{\bar{\alpha}_t} \hat{\mathbf{H}} + \sqrt{1 - \bar{\alpha}_t} \boldsymbol{\epsilon}_t^{(m)}, t \big) \right\|^2. $}
\end{equation}

The second term $\mathbb{E}_{q(\boldsymbol{\pi})}[\log \pi_k]$ involves the Dirichlet distribution $q(\boldsymbol{\pi})$ and has a closed-form solution given by
\begin{equation} \label{eq:Dirichlet_solution}
	\mathbb{E}_{q(\boldsymbol{\pi})}[\log \pi_k] = \psi(\gamma_k) - \psi\bigg( \sum_{j=1}^K \gamma_j \bigg),
\end{equation}
where $\psi(\cdot)$ is the digamma function (defined as $\psi(x) = \frac{d}{dx} \log \Gamma(x)$).

Nevertheless, updating $q(\boldsymbol{z})$ based on \eqref{eq:rho_update} induces an intrinsic bias toward expert models with simpler prior structures. 
Such models tend to be more tolerant to imprecise channel estimates and place higher density to them, particularly when SNRs are low\cite{jakubovitz2019generalization}.
Consequently, we found the algorithm may erroneously favor an oversimplified prior instead of the one that best matches the actual data.
To address this issue, we explicitly incorporate a data consistency term into the update rule. 
Specifically, the refined update rule for $\rho_k$ is formulated as
\begin{equation} \label{eq:refined_qh_update}
	\log \rho_k \propto \log p_k(\hat{\mathbf{H}}) + \mathbb{E}_{q(\boldsymbol{\pi})}[\log \pi_k] + \log p_k(\mathbf{Y}|\hat{\mathbf{H}}_k),
\end{equation}
where $\log p_k(\mathbf{Y}|\hat{\mathbf{H}}_k)$ represents the newly added data consistency term. 
Instead of employing the global MAP estimate $\hat{\mathbf{H}}$ for consistency evaluation, we introduce a refined estimate based on the $k$-th expert, which is denoted as $\hat{\mathbf{H}}_k$.
It is obtained by passing $\hat{\mathbf{H}}$ through the $k$-th DM with a re-noising and denoising process.
To be specific, we corrupt $\hat{\mathbf{H}}$ with a certain level of noise to obtain a noisy state $\tilde{\mathbf{H}}_{\tau}$ via $\tilde{\mathbf{H}}_{\tau} = \sqrt{\bar{\alpha}_\tau}\hat{\mathbf{H}} + \sqrt{1-\bar{\alpha}_\tau}\boldsymbol{\epsilon}_\tau$, where $\tau$ can be directly chosen from $\mathcal{T}_{\text{sub}}$.
A single-step deterministic denoising is then performed using the $k$-th denoising network as
\begin{align}
	\hat{\mathbf{H}}_k = \frac{1}{\sqrt{\bar{\alpha}_\tau}} \left( \tilde{\mathbf{H}}_{\tau} - \sqrt{1-\bar{\alpha}_\tau} \boldsymbol{\epsilon}_{\boldsymbol{\theta}_k}(\tilde{\mathbf{H}}_\tau, \tau) \right).
\end{align} 
The corresponding $\log p_k(\mathbf{Y} | \hat{\mathbf{H}}_k)$ can be evaluated by:
\begin{equation} \label{eq:data_consistency}
	\log p_k(\mathbf{Y} | \hat{\mathbf{H}}_k) = -\frac{1}{2\sigma_n^2} \| \mathbf{Y} - \hat{\mathbf{H}}_k\mathbf{P} \|^2.
\end{equation}  

The underlying intuition for including $\log p_k(\mathbf{Y} | \hat{\mathbf{H}}_k)$ in the update of $\rho_k$ is that if the current estimate $\hat{\mathbf{H}}$ mismatches the prior distribution of the $k$-th expert, the denoising network would fail to effectively predict the injected noise and thus renders the denoised estimate $\hat{\mathbf{H}}_k$ less consistent with the measurement $\mathbf{Y}$.
In other words, the better $\hat{\mathbf{H}}$ aligns with the $k$-th prior, the higher $\log p_k(\mathbf{Y} | \hat{\mathbf{H}}_k)$ tends to be.   

Finally, by aggregating the three terms in \eqref{eq:refined_qh_update}, the variational distribution $q(\mathbf{z})$ is updated by computing its parameters $\boldsymbol{\rho}$ via the following softmax normalization:
\begin{equation} \label{eq:rho_final_update}
	\scalebox{0.97}{$\displaystyle
	\rho_k = \frac{\exp \left( \log p_k(\hat{\mathbf{H}}) + \mathbb{E}_{q(\boldsymbol{\pi})}[\log \pi_k] + \log p_k(\mathbf{Y} | \hat{\mathbf{H}}_k) \right)}{\sum_{j=1}^K \exp \left( \log p_j(\hat{\mathbf{H}}) + \mathbb{E}_{q(\boldsymbol{\pi})}[\log \pi_j] + \log p(\mathbf{Y}|\hat{\mathbf{H}}_j) \right)}. $}
\end{equation}

3) \textbf{Update of} $q(\boldsymbol{\pi})$: The posterior approximation for the mixing coefficients $\boldsymbol{\pi}$ can be derived as:
\begin{equation}
	\begin{aligned} 
		\log q^*(\boldsymbol{\pi}) &\propto \mathbb{E}_{q(\mathbf{z})}[\log p(\mathbf{z} | \boldsymbol{\pi})] + \log p(\boldsymbol{\pi}) \\
		&\propto \mathbb{E}_{q(\mathbf{z})}\left[ \sum_{k=1}^K z_k \log \pi_k \right] + \sum_{k=1}^K (\gamma_k - 1) \log \pi_k \\ 
		&= \sum_{k=1}^K \mathbb{E}_{q(\mathbf{z})}[z_k] \log \pi_k + \sum_{k=1}^K (\gamma_k - 1) \log \pi_k \\
		&= \sum_{k=1}^K (\gamma_k + \rho_k - 1) \log \pi_k.
	\end{aligned}
\end{equation}
Considering $\log \text{Dir}(\boldsymbol{\pi} | \boldsymbol{\gamma}) \propto \sum_{k=1}^K (\gamma_k - 1) \log \pi_k$, it is evident that the optimal variational posterior $q^*(\boldsymbol{\pi})$ follows a Dirichlet distribution $\text{Dir}(\boldsymbol{\pi} | \boldsymbol{\gamma} + \boldsymbol{\rho})$ as:
\begin{equation} \label{eq:gamma_final_update}
	q^*(\boldsymbol{\pi}) = \frac{\Gamma\left(\sum_{k=1}^K (\gamma_k + \rho_k)\right)}{\prod_{k=1}^K \Gamma(\gamma_k + \rho_k)} \prod_{k=1}^K \pi_k^{\gamma_k + \rho_k - 1}.
\end{equation}

Finally, this completes the derivation of the iterative update rules. 
The variational Bayesian inference algorithm proceeds by alternating among the updates of $q(\mathbf{H})$, $q(\mathbf{z})$, and $q(\boldsymbol{\pi})$ until convergence.
For clarity, the proposed algorithm, also referred to as DM-VB is summarized in \textbf{Algorithm}~\ref{alg:DMVB}.
\begin{algorithm}[!t]
	\caption{Channel Estimation via DM-VB}
	\label{alg:DMVB}
	\begin{algorithmic}[1]
		\Require $\mathbf{Y}$, $\mathbf{P}$, $\sigma_n^2$, $s$, pre-trained expert DMs $\{\boldsymbol{\epsilon}_{\boldsymbol{\theta}_k}\}_{k=1}^K$, noise schedule $\{\alpha_t\}_{t=1}^T$, max iterations $L$, tolerance $\eta$.
		
		\State \textbf{Initialize:} $\rho_k^{(0)} = 1/K$, $\gamma_k = 1$ for all $k \in \{1, \dots, K\}$.
		
		\For{$i = 1$ to $L$}
		\State \textbf{Step 1: Update of $q(\mathbf{H})$}
		\State \textbf{Initialize:} $\mathbf{H}_{T} \sim \mathcal{CN}(\mathbf{0}, \mathbf{I})$
		\For{$t = T$ to $1$}
		\State Obtain $\nabla_{\mathbf{H}_t} \log p_t(\mathbf{H}_t)$ and $\nabla_{\mathbf{H}_t} \log p_t(\mathbf{Y}|\mathbf{H}_t)$  \hspace*{\algorithmicindent}\hspace*{\algorithmicindent} based on \eqref{eq:prior_mix_score} and \eqref{eq:likelihood_score}, respectively. 
		\State Obtain $\mathbf{H}_{t-1}$ based on \eqref{eq:h_update}.
		\EndFor
		\State Update the channel estimate $\hat{\mathbf{H}} \leftarrow \mathbf{H}_0$.
		
		\State \textbf{Step 2: Update of $q(\mathbf{z})$}
		\State Obtain $\log p_k(\hat{\mathbf{H}})$, $\mathbb{E}_{q(\boldsymbol{\pi})}[\log \pi_k]$ and $\log p_k(\mathbf{Y} | \hat{\mathbf{H}}_k)$ 
		\hspace*{\algorithmicindent} for each $k$ based on \eqref{eq:prior_approx}, \eqref{eq:Dirichlet_solution} and \eqref{eq:data_consistency}, respectively.   
		\State Update the posterior responsibilities $\boldsymbol{\rho}^{(i)}$ via \eqref{eq:rho_final_update}.
		
		\State \textbf{Step 3: Update of $q(\boldsymbol{\pi})$}
		\State Update the posterior distribution $q(\boldsymbol{\pi})$ via \eqref{eq:gamma_final_update}.
		
		\If{$\|\boldsymbol{\rho}^{(i)} - \boldsymbol{\rho}^{(i-1)}\|_2 < \eta$}
		\State \textbf{break}
		\EndIf
		\EndFor
		
		\State \Return $\hat{\mathbf{H}}$
	\end{algorithmic}
\end{algorithm}

\subsection{Computational Complexity}
The overall computational complexity of the proposed method consists of two aspects, i.e., offline training and online inference.

Regarding the offline training phase, the computational cost is primarily dominated by the training of $K$ expert DMs.
Assuming a training dataset size of $|\mathcal{D}|$ and $N_{ep}$ training epochs used for all experts, the offline training complexity is on the order of $\mathcal{O}(K \cdot N_{ep} \cdot |\mathcal{D}| \cdot \mathcal{C}_{net})$, where $\mathcal{C}_{net}$ denotes the number of floating-point operations (FLOPs) required for a single forward pass of the denoising network.
Since the offline training can be performed in advance, it does not result in any inference latency.

For the online inference phase, the computational complexity mainly comes from the $L$ iterations of the variational updates.
The update of $q(\mathbf{H})$ requires $T$ reverse diffusion steps and thus constitutes the major computational cost.
Each reverse diffusion step involves $K$ forward passes of the expert networks to obtain the prior score, along with a few matrix-vector calculations to compute the likelihood score. 
It can be easily verified that the complexity of updating $q(\mathbf{H})$ scales as $\mathcal{O}(T (K \mathcal{C}_{net} + N_r N_t N_p))$.
The update of $q(\mathbf{z})$ requires evaluating the ELBO and data consistency terms for each expert.
Given the time-subsampling strategy with $|\mathcal{T}_{sub}|$ points, the computational complexity for this is reduced to $\mathcal{O}(K (|\mathcal{T}_{sub}| + 1) \mathcal{C}_{net})$.
Thus, the total computational complexity of our proposed method is on the order of $\mathcal{O}(L \cdot [T (K \mathcal{C}_{net} + N_r N_t N_p) + K |\mathcal{T}_{sub}| \mathcal{C}_{net}])$.

\section{Simulation Results} \label{sec:simulation}
In this section, simulation results are provided to validate the effectiveness and superiority of the proposed DM-empowered variational Bayesian method (DM-VB).

\subsection{Experimental Setup}
Channel datasets used in the simulations are generated according to the 3GPP TR 38.901 CDL channel models.
The CDL family provides a standard representation of realistic multipath propagation environments, capturing diverse spatial and temporal characteristics. 
Specifically, CDL-A, CDL-B and CDL-C are designed to represent non-line-of-sight (NLOS) environments, while CDL-D is constructed for line-of-sight (LOS) scenarios\cite{zhu20213gpp}.
In our simulations, to implement the proposed approach, we train a DM for each CDL model, leading to $K=4$ experts in total.  


\begin{table}[t!]
	\centering
	\caption{Channel Simulation Parameters}
	\normalsize
	\label{tab:CDL_params}
	\begin{tabular}{|l|l|}
		\hline
		Delay Profile & CDL-A, CDL-B, CDL-C, CDL-D \\ \hline
		Antenna Array Type & ULA \\ \hline
		$(N_t, N_r)$ & (64, 16) \\ \hline
		Antenna Spacing & $\lambda/2$ \\ \hline
		Sampling Rate & 15.36 MHz \\ \hline
		Carrier Frequency & 40 GHz \\ \hline
		Delay Spread & 30 ns \\ \hline
		Doppler Shift & 5 Hz \\ \hline
		$N_f$ & 14 \\ \hline
	\end{tabular}
\end{table}


The parameters used for generating CDL channels is presented in Table \ref{tab:CDL_params}. 
For each CDL profile, we generate 10,000 training samples to represent a data-rich scenario, while a subset of 1,000 samples is drawn from this set for the data-scarce case.
Additionally, another $100$ channel samples are generated for testing. 
In our simulations, we use the channel response associated with the first subcarrier as in \cite{arvinte2022mimo}.
The channel samples are normalized such that each entry has unit average power and 
the pilot matrix $\mathbf{P}$ is composed of randomly selected quadrature phase shift keying (QPSK) symbols with unit power, i.e., $\left[1/\sqrt{2}\pm1/\sqrt{2}\text{j}\right]$. 
Therefore, the average SNR can be defined as $N_{t} / \sigma_{n}^2$.

The training and inference are operated on a device equipped with an NVIDIA GeForce RTX 5090 D GPU and an Intel Core Ultra 9 285K CPU.
During training for each CDL profile, the batch size is set to 128, and we train each denoising network with 400 epochs using the Adam optimizer with a fixed learning rate of 0.0001.
The noise parameters $\{\alpha_t\}_{t=1}^T$ are chosen according to the linear schedule in \cite{fesl2024asymptotic} and the total number of steps is set to $T = 100$. 
The channel estimation performance is evaluated by the normalized mean square error (NMSE), which is defined as follows:
\begin{equation}
	\text{NMSE} = \mathbb{E} \left[ \frac{\|\mathbf{H} - \hat{\mathbf{H}}\|_{\mathrm{F}}^2}{\|\mathbf{H}\|_{\mathrm{F}}^2} \right].
\end{equation}

\begin{figure*}[!t]
	\centering
	\includegraphics[width=0.86\linewidth]{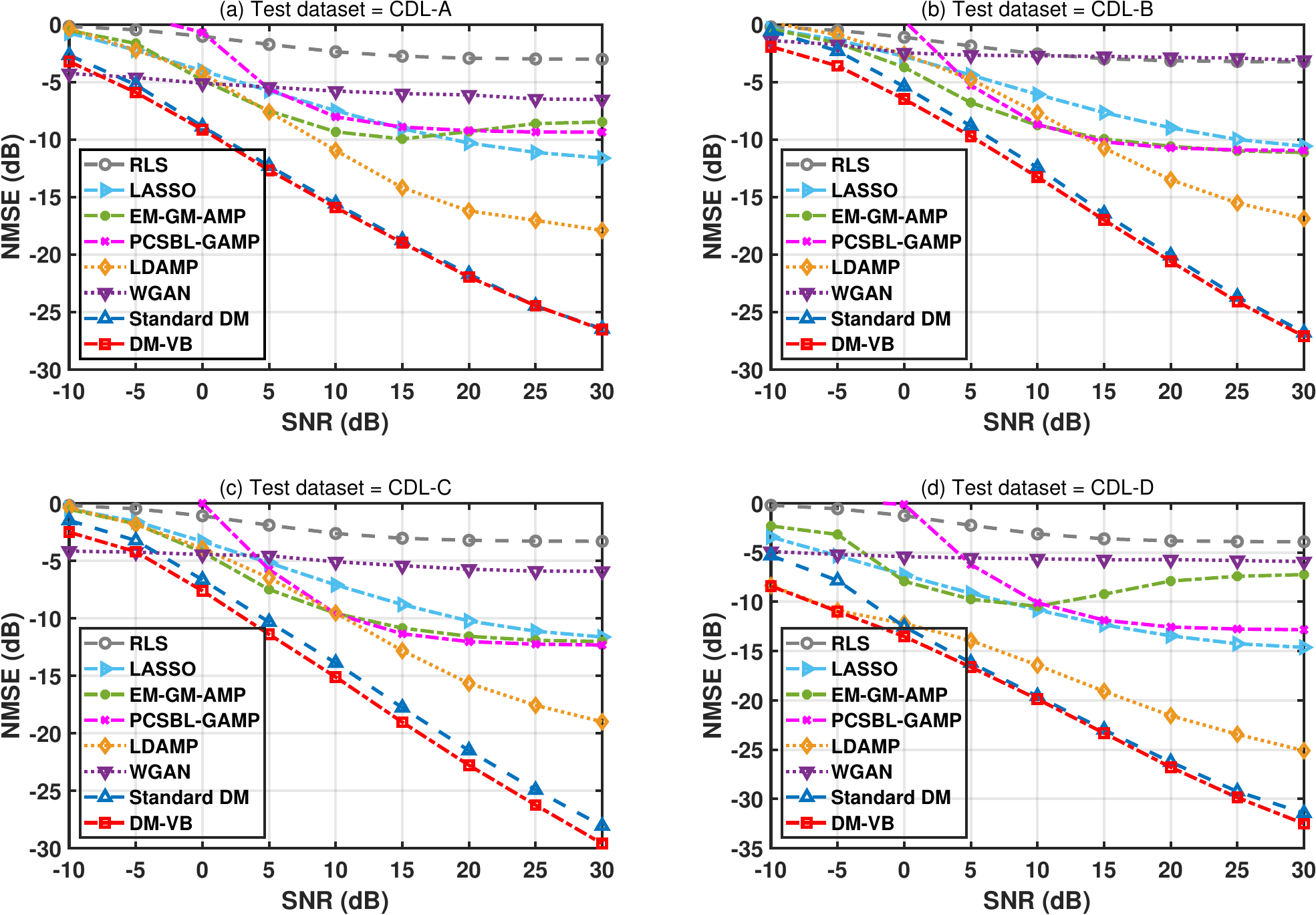}
	\caption{NMSE results among different channel estimation methods under balanced CDL datasets when $\alpha = $ 0.5.}
	\label{fig:balanced_ABCD}
\end{figure*}

\begin{figure*}[!t]
	\centering
	\includegraphics[width=0.9\linewidth]{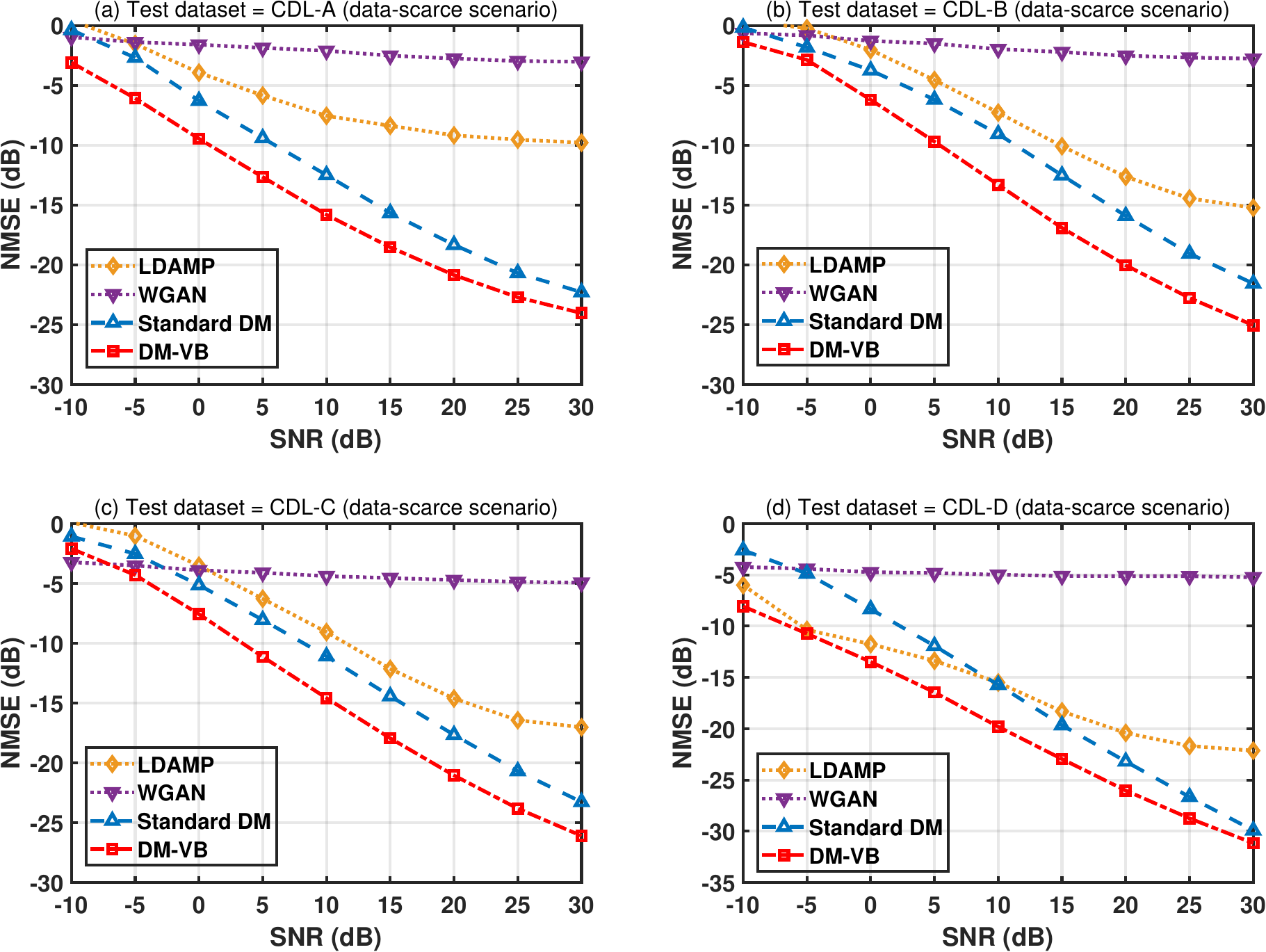}
	\caption{NMSE results among different channel estimation methods under imbalanced CDL datasets when $\alpha = $ 0.5. In each subfigure, the CDL profile designated as the data-scarce scenario is trained using only 1,000 samples, while the remaining CDL profiles utilize the full set of 10,000 channel realizations.}
	\label{fig:imbalanced_ABCD}
\end{figure*}

\subsection{Baselines}
The following channel estimation methods are employed as baselines for comparison with the proposed method, in which the first four methods are model-based, while the remaining three are data-driven approaches.  
\begin{itemize}
	\item 
	\textbf{Regularized LS (RLS)}\cite{rifkin2003regularized}: RLS introduces a regularization term to the standard LS in order to handle the underdetermined regime.
It is given by
	\begin{equation}
		\hat{\mathbf{H}}_{\text{RLS}} = \mathbf{Y}\mathbf{P}^H (\mathbf{P}\mathbf{P}^H + \sigma_n^2 \mathbf{I})^{-1}. 
	\end{equation}
	
	\item 
	\textbf{LASSO}\cite{tibshirani1996regression}: A CS-based channel recovery method that leverages the channel sparsity in the angular domain via the $l_1$-norm regularized optimization problem, i.e.,
	\begin{equation}
		\hat{\mathbf{H}}_{\text{LASSO}} = \underset{\mathbf{H}}{\operatorname{argmin}} \  \frac{1}{2}\|\mathbf{Y} - \mathbf{H}\mathbf{P}\|_\mathrm{F}^2 + \lambda \|\mathbf{F}_r^H \mathbf{H} \mathbf{F}_t\|_{\mathrm{m}_1},
	\end{equation}
	where $\lambda$ is the regularization parameter and $\mathbf{F}_r, \mathbf{F}_t$ denote the Discrete Fourier Transform (DFT) matrices of size $N_r \times N_r$ and $N_t \times N_t$, respectively.
	
	\item 
	\textbf{EM-GM-AMP}\cite{vila2013expectation}: This CS-based approach models the sparse angular domain channel using a Gaussian mixture (GM) prior. 
	It employs the Expectation-Maximization (EM) algorithm to learn the hyperparameters, which are then utilized by the AMP framework for robust channel estimation.
	
	\item
	\textbf{PCSBL-GAMP}\cite{fang2016two}: This approach employs a 2D pattern-coupled hierarchical Gaussian prior to exploit the channel's block-sparse structure without prior knowledge of its cluster patterns.
	For efficient Bayesian inference, it leverages generalized AMP within an EM framework.
	
	\item 
	\textbf{LDAMP}\cite{he2018deep}: A deep learning–based estimator that unfolds the AMP algorithm by replacing its linear shrinkage operator with a CNN denoiser. 
	
	\item 
	\textbf{Wasserstein GAN (WGAN)}\cite{balevi2021high}: WGAN provides an unsupervised generative framework that exploits the channel distribution via adversarial training. 
	At inference, it estimates the channel by optimizing the latent input to the pre-trained generator to minimize the discrepancy between observed and estimated measurements.
	
	\item 
	\textbf{Standard DM-Based Method}: As in \cite{zhou2025generative}, it trains a single DM as the prior, and then leverages the deterministic posterior sampling-based approach to perform the channel estimation. 	
\end{itemize}

Note that the true type of channel distribution from which the channel is generated is unknown during inference. Therefore, all data-driven baselines are trained on a composite dataset that aggregates all the channel samples from the four CDL scenarios (CDL-A to CDL-D). 
For instance, the standard DM-based method learns a single DM as the universal prior by training on this mixed dataset. 
Since the standard DM-based method employs the same CNN architecture as the one for DM-VB, to ensure a fair comparison, the total number of network parameters used in the standard DM is set to approximately $K$ times that of each expert DM in our framework.
Hyperparameters of all competing algorithms are tuned to achieve favorable performance.

\subsection{Estimation Performance}
We evaluate the performance of the proposed DM-VB approach against the baseline algorithms on different CDL channel profiles. 
The first simulation is conducted under a balanced data setting, using all 10,000 channel samples from each CDL profile for training.
Accordingly, the aggregated dataset includes 40,000 samples.
Fig. \ref{fig:balanced_ABCD} depicts the NMSEs of respective methods as the SNR varies from -10 dB to 30 dB, where the pilot density is set to $\alpha = 0.5$ ($N_p = 32$).
We see that the proposed method outperforms RLS, all CS-based methods, LDAMP and WGAN by a wide margin across different CDL models, and it consistently achieves better estimation accuracy than the standard DM-based method. 
In particular, classical LS estimators such as RLS fail to yield reliable channel estimates in the underdetermined regime.
Compared with RLS, CS-based methods including LASSO, EM-GM-AMP and PCSBL-GAMP, achieve a performance improvement by exploiting the sparse structure inherent in wireless channels. Nevertheless, their estimation performance saturates rapidly as SNR increases.
This is because the practical 3GPP CDL channels contain rich structural information beyond sparsity, which is unable to be utilized by CS-based methods. 
We also observe that WGAN achieves the poorest performance among data-driven approaches. 
As noted in \cite{arvinte2022mimo,bora2017compressed}, this unsatisfactory behavior is probably attributed to the mode collapse and suboptimal inversion of the latent representation by standard optimizers such as Adam. 
In contrast, LDAMP and the standard DM demonstrate competitive performance.

Particularly, the standard DM-based method, which is trained on the mixed dataset, achieves the best performance among all baselines and closely approaches that of the proposed method.
The reason lies in that although 3GPP CDL models are categorized into distinct scenarios, they still share inherent similarities in their spatial-temporal structures.
This makes it possible for the standard DM method to effectively learn common features of the wireless channel, thus resulting in a competitive generalization performance.
We observe that our proposed method consistently outperforms the standard DM method by a slight margin.
This advantage arises from the MoE framework, which enables the proposed method to more accurately capture the essential channel features used for channel estimation.

\begin{figure*}[!t]
	\centering
	\includegraphics[width=0.83\linewidth]{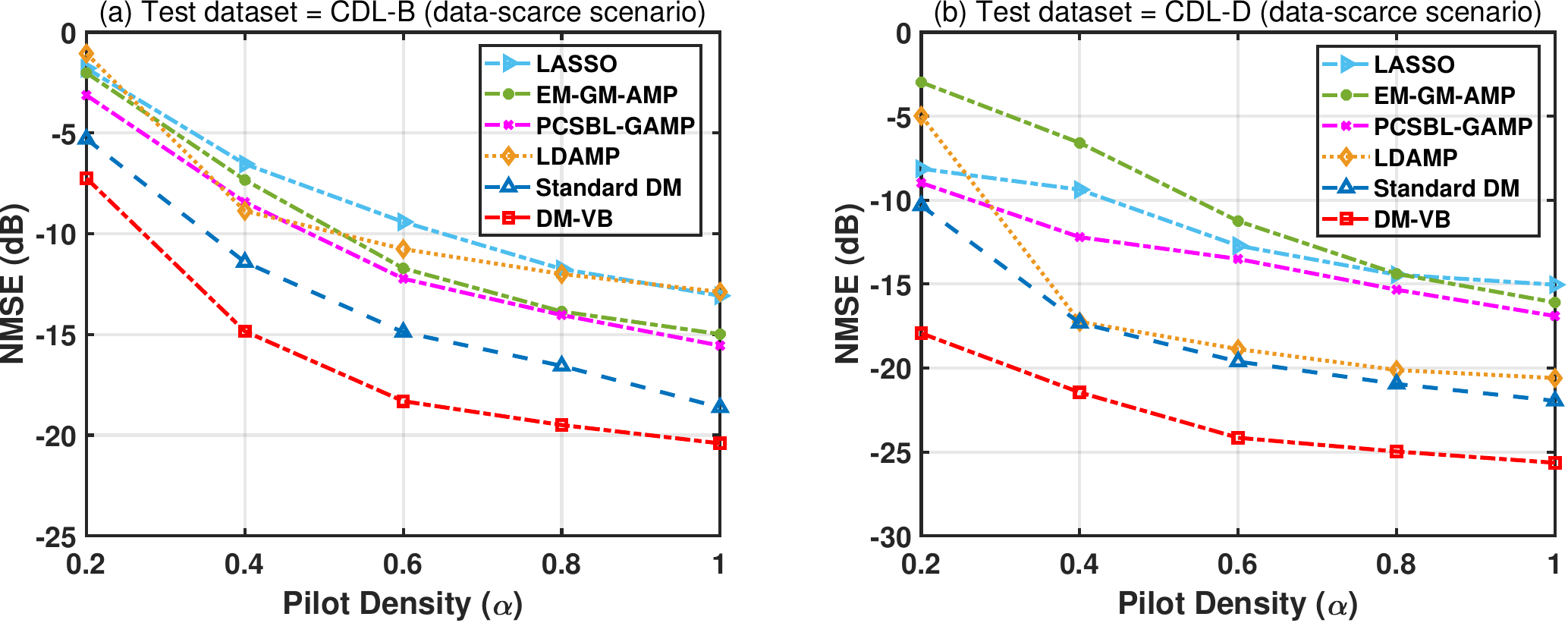}
	\caption{NMSE results among different channel estimation methods with varying pilot density $\alpha$ under imbalanced CDL datasets when SNR $=$ 15 dB.}
	\label{fig:alpha_BD}
\end{figure*}

To more clearly highlight the superiority of our approach, we evaluate the performance of respective methods under an imbalanced training data setting.
Specifically, among the four CDL profiles, one is designated as a data-scarce scenario with only 1,000 training samples, while for the remaining three profiles, each still retains the full set of 10,000 samples for training. 
The three data-driven baselines, namely, the LDAMP, WGAN, and the standard DM, are trained based on the aggregated data set containing a total of 31,000 channel samples. 
For our DM-VB method, a dedicated DM is trained for each CDL profile using 10,000 channel samples for those data-rich CDL profiles and 1,000 channel samples for the data-scarce one. 
Under this imbalanced training data setup, the standard DM method, along with the other two data-driven baselines, is expected to suffer severe performance degradation on the data-scarce CDL profile, since the trained network tends to prioritize the channel features associated with those data-rich CDL profiles and ignore features of the profile with limited data samples. 
In contrast, the proposed DM-VB method can effectively circumvent this issue via the automatic model adaptation mechanism. 

To show this, Fig.~\ref{fig:imbalanced_ABCD} provides the estimation performance of respective methods evaluated on the data-scarce CDL profile.       
It can be observed that the proposed DM-VB method significantly outperforms other baselines.
Particularly, for the CDL-D model characterized by distinctive LOS features, our method achieves over 5 dB performance improvement over the standard DM method.
Similarly, when the CDL-B profile is selected as the data-scarce scenario, the proposed approach outperforms the standard DM by nearly 4 dB.
Consistent performance gains are also observed in the CDL-A and CDL-C scenarios, verifying the superiority of DM-VB.  
The results demonstrate that, by assigning dedicated DM experts to their corresponding propagation environments, DM-VB effectively addresses the problem of feature dilution induced by imbalanced training data.

We further investigate the performance of the proposed method under different pilot densities $\alpha$.
Fig. \ref{fig:alpha_BD} illustrates the NMSE performance of different estimators as a function of $\alpha$ at a fixed $\text{SNR}=15$ dB.
We focus on the two cases in which CDL-B and CDL-D are respectively designated as the data-scarce scenarios, and we omit the results of RLS and WGAN from this investigation due to their underwhelming performance.
We see that the proposed DM-VB method consistently yields the highest estimation accuracy compared with other baselines.
Notably, when CDL-D is set as the data-scarce scenario, the proposed method achieves a remarkable performance improvement of approximately 7.5 dB over the standard DM and 13 dB over LDAMP in the low pilot density regime, i.e., $\alpha=0.2$.
For CDL-B, our method also yields gains of roughly 3.5 dB.
The results highlight that, rather than diluting environment-specific structures as seen in the standard DM method, our method is able to preserve the distinct structural features of each channel type via the automatic model adaption mechanism. 
This enables informative guidance for robust channel recovery in the limited measurement regime.

\subsection{Convergence and Complexity}
We first investigate the convergence behavior of the proposed method. Specifically, both the convergence of the loss function during the training of individual expert DMs, and the convergence of the Bayesian inference algorithm developed for channel estimation are examined.
In addition, we compare the DM-VB method with the benchmark approaches in terms of number of parameters, FLOPs, and latency.

\begin{figure}[t]
	\centering
	\begin{subfigure}[b]{0.4\textwidth}
		\centering
		\includegraphics[width=\textwidth]{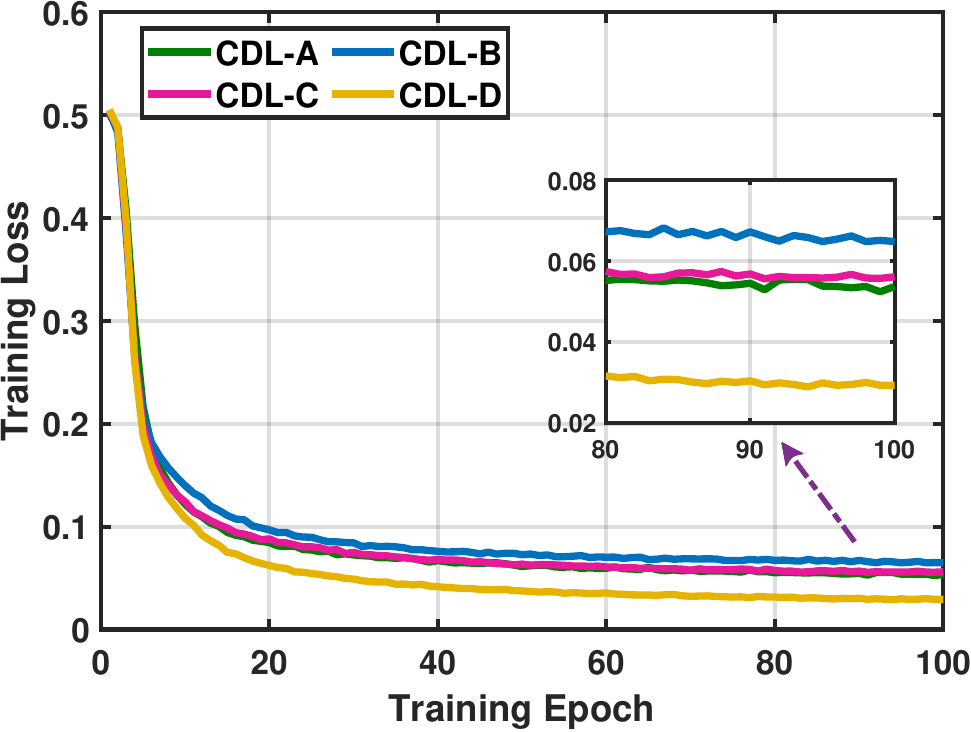}
		\caption{}
		\label{fig:training_loss}
	\end{subfigure}
	
	
	\begin{subfigure}[b]{0.4\textwidth}
		\centering
		\includegraphics[width=\textwidth]{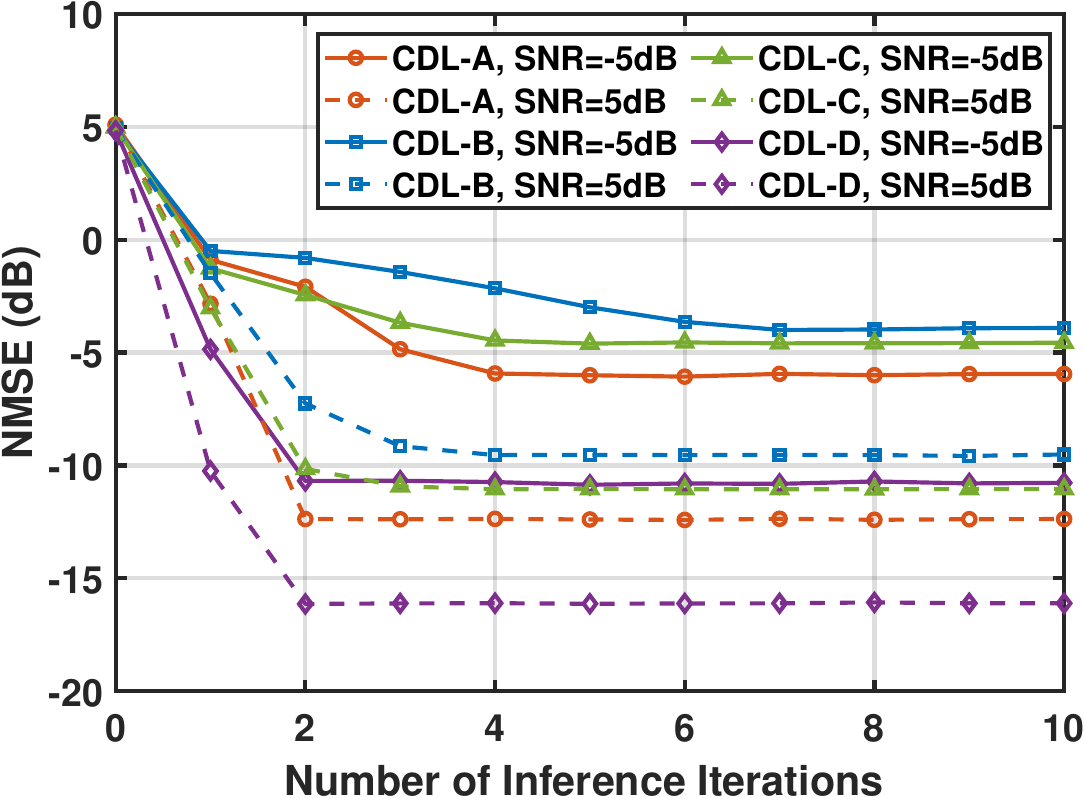}
		\caption{}
		\label{fig:test_nmse}
	\end{subfigure}
	
	\caption{Convergence behavior of the proposed method: (a) Training convergence; (b) Inference convergence.}
	\label{fig:convergence}
\end{figure}

Fig. \ref{fig:training_loss} depicts the DM's training loss versus the number of epochs trained on different CDL datasets. 
For clarity, only the training loss of the first 100 epochs is shown. 
As observed, the DM exhibits stable convergence across all CDL channel distributions, with rapid loss reduction in the early stages of training.
Fig. \ref{fig:test_nmse} illustrates the convergence behavior of the proposed method in terms of NMSE versus the number of outer iterations $L$ during the inference phase.  
We evaluate the performance across different CDL channel models with the pilot density set to $\alpha = 0.5$.
It is evident that the DM-VB algorithm converges rapidly under all scenarios, with the NMSE improving significantly in the first few iterations.
Moreover, a higher SNR generally leads to faster and more stable convergence.

Table~\ref{tab:complexity} presents the computational complexity of EM-GM-AMP, LDAMP, WGAN, the standard DM and the proposed method. 
The number of FLOPs and the latency are measured at $\alpha = 0.5$ and averaged over 100 channel realizations.
For our proposed method, the number of iterations required for convergence varies with SNR, as shown in Fig.~\ref{fig:test_nmse}. Thus, we report number of per-iteration FLOPs and per-iteration latency for our proposed method in Table~\ref{tab:complexity}. 
We observe that the proposed method achieves notably lower latency than EM-GM-AMP and WGAN.
Its per‑iteration cost is higher than that of the standard DM and LDAMP, due to the use of $K = 4$ expert DMs whose forward passes are required in each variational update.
Nevertheless, the lightweight architecture of each expert, along with efficient GPU parallelization, enables its actual latency per iteration to remain comparable to the standard DM.
Moreover, DM-VB converges rapidly in practice (typically within $5$ iterations), which ensures the total amount of inference time acceptable.
Overall, considering the substantial performance advantage over the state-of-the-art methods, our proposed method offers a favorable trade‑off between inference latency and estimation accuracy.

\begin{table}[t]
	\centering
	\caption{Complexity comparison of different methods.}
	\small
	\label{tab:complexity}
	\begin{tabular}{|l|c|c|c|}
		\hline
		\textbf{Method} & \textbf{Parameters} & \textbf{FLOPs} & \textbf{Latency [ms]} \\
		\hline
		EM-GM-AMP & - & $\mathcal{O}(N_tN_r)$ & $180.74$ \\ 
		LDAMP     & $4.8 \times 10^{6}$ & $6.3$G & $4.86$ \\
		WGAN      & $1.6 \times 10^{6}$ & $273$G & $140.65$ \\
		Standard DM   & $5.5 \times 10^{4}$ & $10.9$G & $3.76$ \\
		DM-VB  & $4 \times 1.4 \times 10^{4}$ & $12.1$G & $4.61$\\
		\hline
	\end{tabular}
\end{table}

\section{Conclusion} \label{sec:conclusion}
In this paper, we proposed a novel DM framework integrated with variational Bayesian inference to enable automatic model adaptation for MIMO channel estimation.
The framework leverages multiple pre-trained DMs to capture the distinct distribution of each channel type, and then infers the weights of expert assignment for the current channel via an iterative variational procedure.
In contrast to conventional DM-based methods that rely on a single trained prior, the MoE architecture of our proposed method enables a more accurate prior, particularly under imbalanced training data scenarios.
Simulation results demonstrate that the proposed method achieves a significant performance improvement over all benchmarks trained on aggregated datasets.
In addition, owing to the lightweight network architecture and the fast convergence of the inference algorithm, our approach guarantees competitive estimation latency, striking an effective balance between complexity and accuracy.

\bibliography{ref}
\bibliographystyle{IEEEtran}
\end{document}